\documentclass[aps,prl,twocolumn,superscriptaddress]{revtex4-2}

\usepackage{graphicx}
\usepackage{dcolumn}
\usepackage{bm}
\usepackage{hyperref}
\usepackage{units}
\usepackage{amsmath, amsfonts}
\usepackage{ulem}

\begin{document}


\title{Critical clusters in liquid crystals: Fractal geometry and conformal invariance}


\author{Renan A. L. Almeida}
\email[]{renan.almeida@ufrgs.br}
\email[]{ra.lisboaalmeida@gmail.com}
\affiliation{Instituto de F\'\i sica, Universidade Federal do
Rio Grande do Sul, CP 15051, 91501-970, Porto Alegre RS, Brazil}

\author{Jeferson J. Arenzon}
\email[]{arenzon@if.ufrgs.br}
\affiliation{Instituto de F\'\i sica, Universidade Federal do
Rio Grande do Sul, CP 15051, 91501-970, Porto Alegre RS, Brazil}
\affiliation{Instituto Nacional de Ciência e Tecnologia - Sistemas
Complexos, Rio de Janeiro RJ, Brazil}


\date{\today}

\begin{abstract}
We study the two-dimensional domain morphology of twisted nematic liquid crystals during their phase-ordering kinetics 
[R. A. L. Almeida, Phys. Rev. Lett. \textbf{131}, 268101 (2023)], 
which is a physical candidate to self-generate critical clusters in the percolation universality class. 
Here we present experimental evidence that large clusters and their hulls are indeed both fractals with dimensions 
of the corresponding figures in critical percolation models.
The asymptotic decay of a crossing probability, from a region in the vicinity of the origin to the boundary of disks, is described by the Lawler-Schramm-Werner theorem provided that a microscopic length in the original formulation is replaced by the coarsening length of the liquid crystal.
Furthermore,
the behavior for the winding angle of large loops is, at certain scales, 
compatible with that of Schramm-Loewner evolution curves with diffusivity $\kappa = 6$. 
These results show an experimental realization of critical clusters in phase ordering.

\end{abstract}


\maketitle




Phase-ordering kinetics~\cite{Bray02, Cugliandolo15} 
in two-dimensional scalar systems endowed with a $\mathbb{Z}_2$ symmetry
can be found in several branches of science.
In good approximation, 
it is observed in real systems like liquid crystals \cite{Sicilia08, Almeida23},
spatial segregation of dense bacterial populations~\cite{Mcnally17},
and, arguably, in vortex patterns of collectively moving cells~\cite{Andersen24}.
Simple models for the kinetics of phase transitions in classical~\cite{Arenzon07} 
and quantum matter~\cite{Hofmann14},
including models for opinion formation~\cite{Dornic01,Latoski24},
are also examples in the list.

\begin{figure}[!t]
\includegraphics[width=0.23\textwidth]{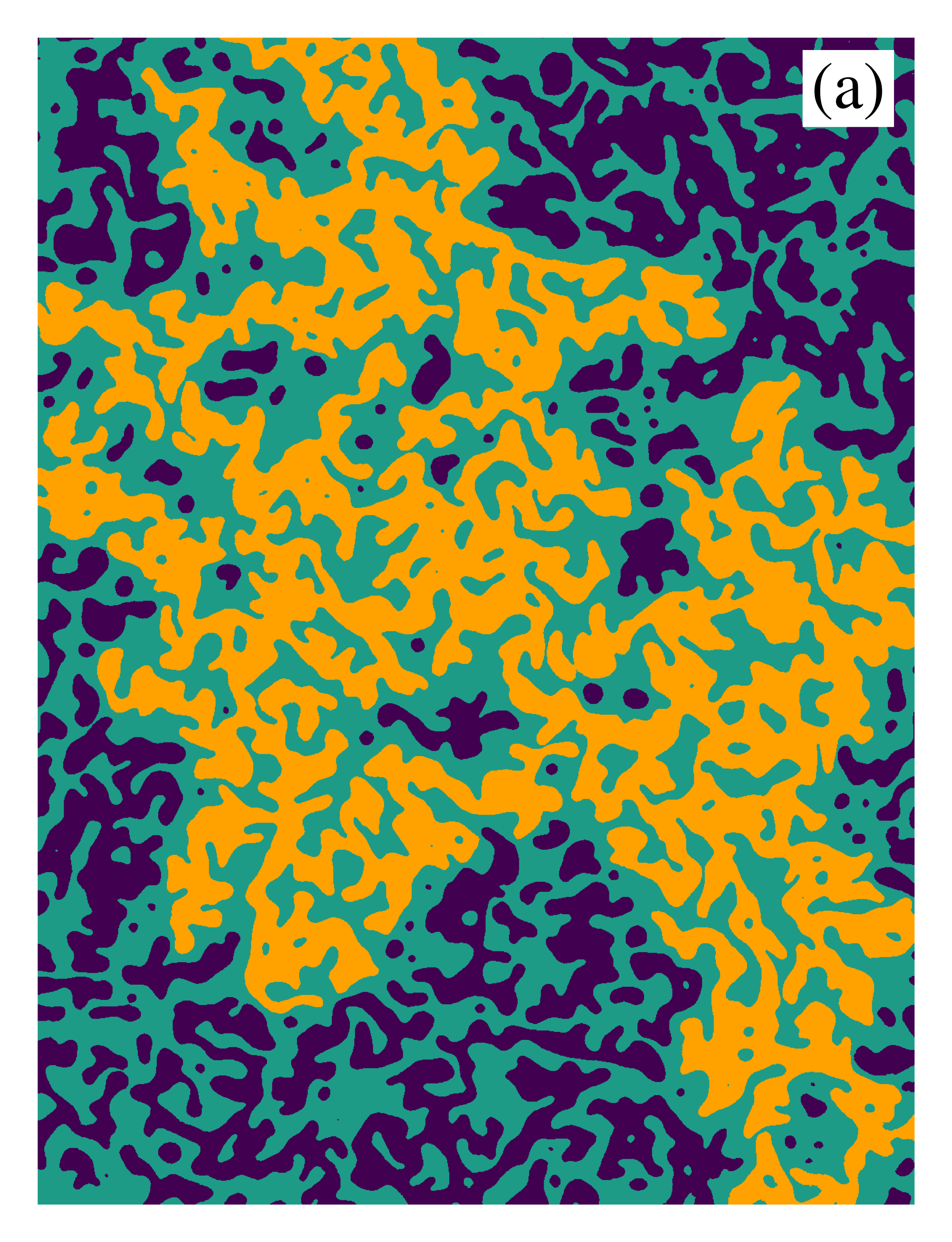}
\includegraphics[width=0.23\textwidth]{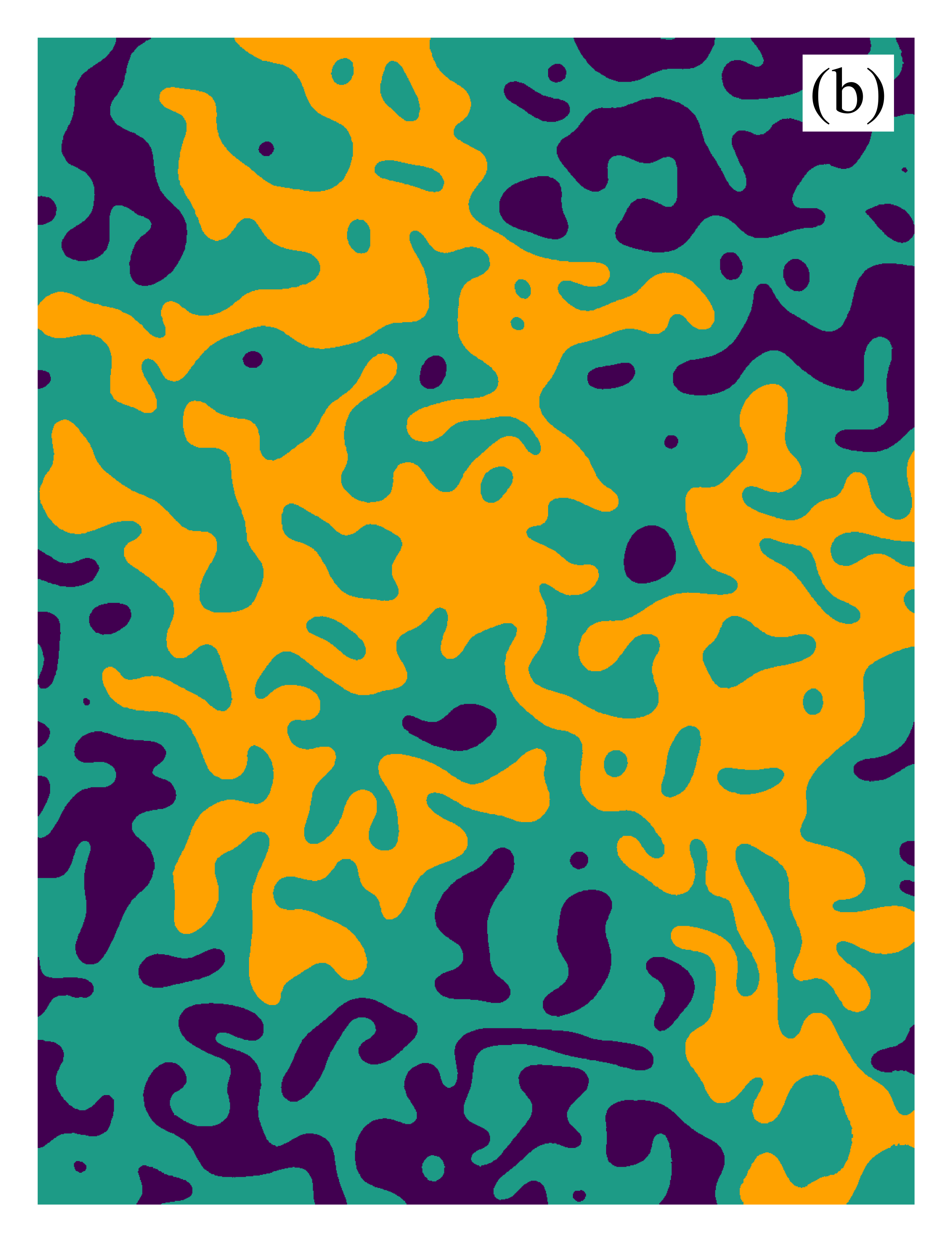}
\includegraphics[width=0.23\textwidth]{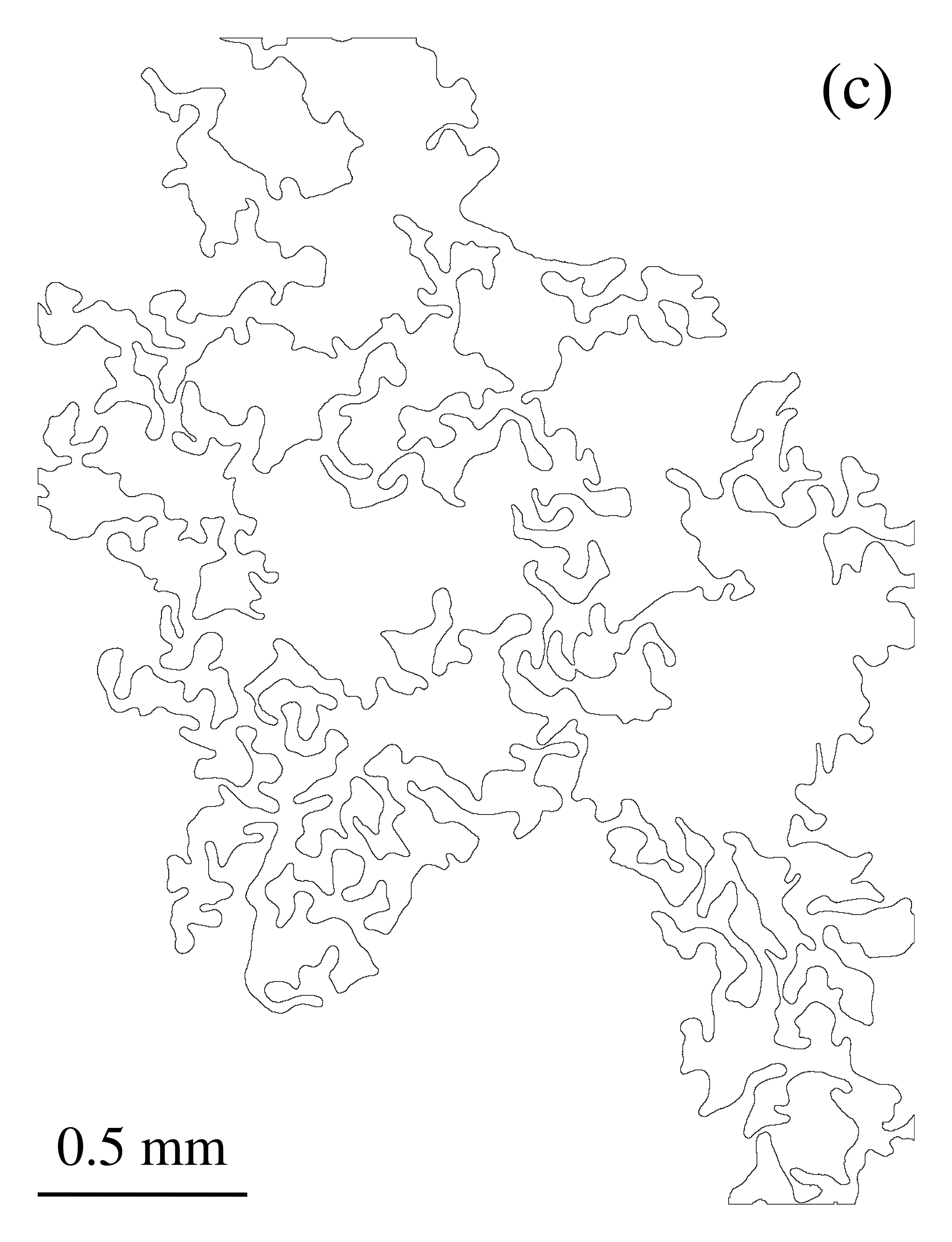}
\includegraphics[width=0.23\textwidth]{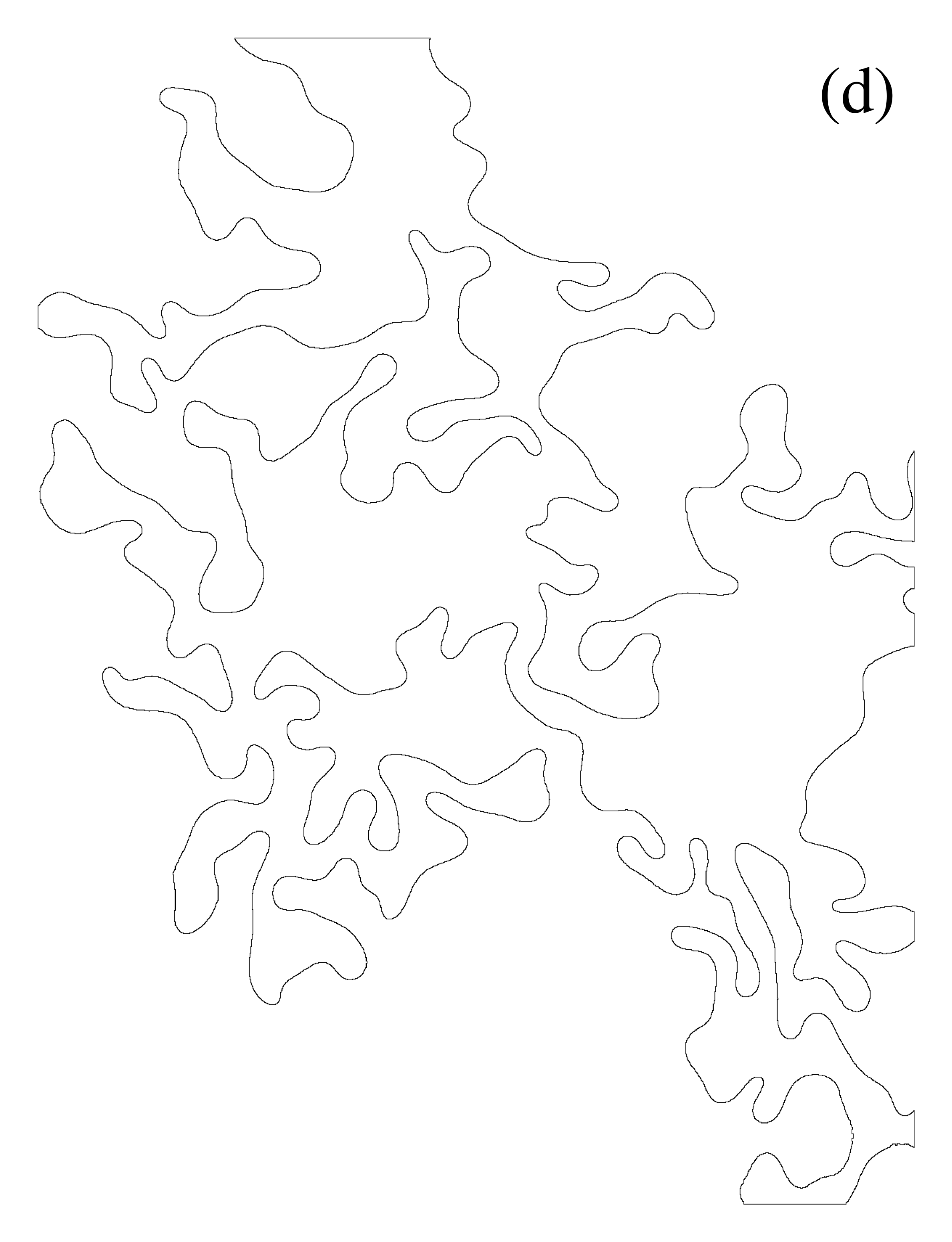}
\caption{\label{fig_cluster}
Example of domain morphology developed in the ordering kinetics of twisted nematic phases at $\unit[2.0]{s}$ (left) and $\unit[8.0]{s}$ (right) after the quench.
The two competing phases are depicted in violet and green colors, (a) and (b).
The largest cluster is highlighted by the orange color.
%
The hull of the largest cluster is shown, (c) and (d).
}
\end{figure}

A well-known characteristic among coarsening systems is that they often develop an intricate domain morphology, e.g. Fig.~\ref{fig_cluster}(a).
The statistical description of such morphology becomes time invariant in the hydrodynamic limit 
when lengths are normalized by an emergent growing scale~---~a statement known as the dynamical scaling hypothesis~\cite{Bray02}.
Conservation laws and symmetries constrain the forms of correlation functions,
thereby signaling the underlying principles by which microscopically disparate systems
are classified into dynamical universality classes~\cite{Bray02, Dornic01}.


A recently recognized feature~\cite{Arenzon07, Barros09, Olejarz12}
is the superuniversal geometry of large clusters
in the percolation class~\cite{Stauffer_book,Araujo14,Saberi15}.
Evolving from random initial conditions, 
the morphology of large systems seems disordered at scales larger than the average domain size,
so that microscopic details are eventually washed out
while the coarse-grained picture becomes similar to a percolation problem in the continuum.
Self-tuning to the exact percolation threshold (1/2) is guaranteed by the $\mathbb{Z}_2$ symmetry.
Notably, the ordering dynamics of finite-size systems quickly converge 
to stable critical configurations 
in a timescale $t_{\textit{p}}$~\cite{Blanchard14} 
that is shorter than that expected from a growth law~\cite{Azevedo22}.
Being superuniversal,
signatures of critical percolation (CP) are observed in ordering model systems regardless of conservation laws~\cite{Sicilia07, Sicilia09},
absence of surface tension~\cite{Tartaglia18},
and, 
at least for kinetic Ising models with a nonconserved order parameter, 
they also survive amid
quenched disorder~\cite{Sicilia08_disorder, Corberi17},
slow cooling protocols~\cite{Ricateau18},
long-range interactions~\cite{Agrawal22},
and geometric frustration~\cite{Agrawal23}.


There are several important consequences of CP in phase ordering.
First, analytical results were obtained for the evolution of geometric quantities in classes of nonconserved~\cite{Arenzon07, Sicilia07} and conserved fields~\cite{Sicilia09}. 
Second, 
the crossing types of percolating domains in finite geometries can be anticipated from $t_{\textit{p}}$ 
by a beautiful application of the crossing probability formulas derived for the geometric problem~\cite{Barros09,Olejarz12, Tartaglia16, Corberi17}.
Third,
it indicates that clusters produced in these nonequilibrium systems 
may be characterized by a set of nontrivial fractal dimensions,
some of which are related to standard scaling exponents~\cite{Stauffer_book,Araujo14,Saberi15}.
Moreover, the hulls should be designed as conformally invariant objects, 
whose description in the continuum limit may be given by the family of Schramm-Loewner evolution (SLE) curves~\cite{Schramm00,Cardy05,Bauer06} with diffusivity $\kappa = 6$~\cite{Smirnov01}.
%
%
%
Albeit the first and second cases above have been experimentally confirmed~\cite{Sicilia08, Almeida23},
the third one has not been settled.
The question is whether a complex system 
would share the same fractality, 
percolation exponents, 
and conformal invariant behavior of oversimplified lattice models.


To probe that question, we study the experimental domain morphology of twisted nematic liquid crystals~\cite{Almeida23}.
The system is chosen in view of its large statistics, high accuracy, and known timescales.
In this work, we measure the fractal dimensions of large domains and hulls. 
A theorem for the crossing probability of clusters in a circular geometry is translated to and tested in the nonequilibrium system.
By evaluating an expected behavior for conformally invariant curves described by SLE, 
we extract the underlying diffusivity $\kappa$ from the experimental data.
The time dependence of the analyzed observables are rationalized by the dynamical scaling hypothesis.
Our results show that critical clusters in the percolation universality class are self-generated during the ordering.
Below, we describe the experimental methods. 
Analysis and discussion of the results are given in sequence.



\textit{Experiments}.---
One of us used an electroconvection cell~\cite{Almeida21} of parallel plates containing
N-4-methoxybenzylidene-4-butylaniline (purity $>\unit[98]{\%}$) with $\unit[0.01]{wt}$ of tetrabutylammonium bromide.
The material was confined in a volume, $\unit[16]{mm} \times \unit[16]{mm} \times \unit[12]{\mu m}$, 
delimited by polyester spacers and rubbed polyvinyl alcohol layers.
Boundary conditions set by these layers allowed the nematic field to adopt left- or right-hand twists.


The cell was subjected to an a.c. electric field (\unit[70]{V}; \unit[100]{Hz}) to induce the Dynamical Scattering Mode 2 (DSM2)~\cite{Joetz86, Kai89}. 
DSM2 plays the role of a disordered-like initial condition because of the short nematic correlations in space and time ($\sim \unit[1]{\mu m}$; \unit[10]{ms})~\cite{Joetz86}.
After driving DSM2 for \unit[2]{min}, 
the external field was suddenly turned off~---~definition of $t = \unit[0]{s}$. 
This triggered the ordering of antagonistic twisted nematic phases that we are interested in.


To distinguish between the phases, the cell was illuminated with circularly polarized green-filtered light using an Olympus IX73 inverted microscope.
Light patterns transmitted through the material and the objective lens
were recorded by a charge-coupled device camera 
as images of area $(L_{\textrm{x}} \times L_{\textrm{y}}) = (1208a \times 1608a) = (\unit[2.2]{mm} \times \unit[2.9]{mm})$ arranged in a squared mesh of pixel length $a = \unit[1.82]{\mu m}$.
The dynamics were recorded at $\unit[5]{s^{-1}}$ frame rate over \unit[30]{s}.
In total, $10^3$ independent ordering histories were collected~\cite{Almeida23}.
The cell temperature was kept at $\unit[25]{^{\circ}C}$ with fluctuations of a few tens of \unit{}{mK}. 
More details of the setup are available~\cite{Almeida21}.


In the image processing step~\cite{Almeida23}, 
a binary variable $s(\bm{r},t)$ was assigned to each pixel: 
$s = -1$ for a twisted phase; and $s = +1$ for the counterpart phase.
In the analysis, neighboring sites are adjacent pixels along the lines or columns of the panels.
A domain (cluster) is a connected path of neighboring sites having the same phase.
A hull is the set of connected neighboring sites on the external boundary of a domain.
Domains and hulls were detected by a labeling~\cite{Hoshen76} and a biased-walker algorithm~\cite{Sicilia07}.


\textit{Typical scales}.---
At a particular timescale $t_{\textrm{p}} = \unit[1.4(1)]{s}$, 
crossing probabilities of large clusters in the liquid crystal follow the probabilities exactly derived for critical percolation models in rectangular geometries. The agreement occurs for rectangles of several aspect ratios (0.2, 0.3, ..., 1) with their larger side $\sim 10^3a$ \cite{Almeida23}.
The ordering proceeds by a curvature-driven motion for interfaces separating the phases; 
the typical domain size grows as $\sim \sqrt{t}$~\cite{Almeida21}. 
Hull-enclosed areas of nearly circular clusters immersed in a sea of the opposite phase shrink at a rate proportional to $\lambda_{\textrm{h}} \approx \unit[767]{\mu m^2 s^{-1}}$~\cite{Almeida21, Almeida23}.
This rate is a good approximation for the changing in area per unit time of domains, 
$\lambda_{\textrm{d}} \approx \lambda_{\textrm{h}}$, 
shown to hold analytically~\cite{Sicilia07}.
The characteristic length for domains and hulls can be defined~\cite{Sicilia07} as $\sqrt{\lambda_{\textrm{d,h}}t}$, 
for $t \gg t_{\textrm{p}}$.



\textit{Fractal dimensions}.---
From $t_{\textrm{p}}$ onwards,  
the largest percolating cluster becomes stable and firmly establishes on the morphology~\cite{Blanchard14}.
Having many holes (i.e., inner domains of the opposite phase), Fig.~\ref{fig_cluster}(a), 
and an irregular contour that is almost self-touching in several places, Fig.~\ref{fig_cluster}(c),
the fractal features of the largest cluster are to be contrasted with the compact shapes and smooth boundaries of small domains.
As time elapses, domain areas coarsen by the shrinking of inner clusters, Fig.~\ref{fig_cluster}(b), 
while winding boundaries are reshaped by flatter segments, Fig.~\ref{fig_cluster}(d).


To quantify the dimensions $D_{\textrm{c}}$ and $D_{\textrm{h}}$ of the largest cluster and its hull, 
we compute the average mass, $M_{\textrm{c}}(L,t)$ and $M_{\textrm{h}}$, of each object in a square patch of size $L \in [a, L_{\textrm{x}}]$.
Sides of the patch are kept parallel to the short and long sides of the panels.
The probing square glides over all possible spatial configurations, while it records the number of pixels (viz., the mass) of the target figure inside the specified region.
Averaging over patches containing some mass and repeating the procedure for different sizes, we obtain a mass curve.
The quantities $M_{\textrm{c}}$ and $M_{\textrm{h}}$ are obtained after performing averages over the ensemble of mass curves at $t$.

\begin{figure}[!t]
\includegraphics[width=0.49\columnwidth]{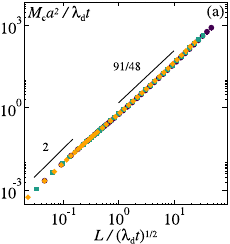}
\includegraphics[width=0.489\columnwidth]{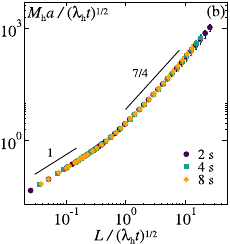}

\caption{\label{fig_massscaling}
Master scaling forms for the averaged mass of the largest domain (a) and hull (b) inside a square patch of side $L$.
In the normalization, $\lambda_{\textrm{h}} = \unit[767]{\mu m^{2}s^{-1}}$; we take $\lambda_{\textrm{d}} = \lambda_{\textrm{h}}$. 
Solid lines are guide to eyes; the values $91/48$ and $7/4$ are the fractal dimensions of clusters and hulls in critical percolation, respectively.
Plots have the same key. 
Uncertainties are one standard deviation.
}
\end{figure}


Because the largest domain progressively gets fatter after $t_{\textrm{p}}$, 
its mass $M_{\textrm{c}}(L,t)$ increases in time for large $L$ [Fig. S1(a), Suppl. Mat.].
Taking the growing lengthscale into account, 
the curves for different times collapse on a master scaling form, Fig.~\ref{fig_massscaling}(a).
This form reads $M_{\textrm{c}}(L,t)a^2/\lambda_{\textrm{d}}t \simeq f(L/\sqrt{\lambda_{\textrm{d}}t})$.
The scaling function behaves as $f(x) \sim x^2$ in $0.02 < x < 0.2$, 
grows slower in the crossover region close to $x = 0.7$, 
and has the algebraic law $f(x) \sim x^{D_{\textrm{c}}}$ at $0.7 \ll x \ll L_{\textrm{x}}/\sqrt{\lambda_{\textrm{d}}t}$. 
The exponent is
consistent with $D_{\textrm{c}} = 91/48 = 1.89...$~\cite{Stauffer_book},
the exact result for the fractal dimension of clusters in percolation theory.
At larger scales there are boundary effects that may affect the results.


%
%
%
%


A similar behavior occurs for the average mass of the largest hull. 
Because the length of a hull tends to diminish despite a thermal roughening, 
$M_{\textrm{h}}(L,t)$ decreases in time for large $L$ 
[Fig. S1(b), Suppl. Mat.].
A scaling relation for this case is $M_{\textrm{h}}(L,t)a/\sqrt{\lambda_{\textrm{h}}t} \simeq g(L/\sqrt{\lambda_{\textrm{h}}t})$, Fig.~\ref{fig_massscaling}(b).
Far from the crossover regime, $g(x') \sim x'$ for $0.02 < x' < 0.2$,
and $g(x') \sim x'^{D_{\textrm{h}}}$ for $0.7 \ll x' \ll L_{\textrm{x}}/\sqrt{\lambda_{\textrm{h}}t}$.
Our results are in agreement with the fractal dimension for hulls in critical percolation~\cite{Saleur87}, $D_{\textrm{h}} = 7/4$, covering more than one decade in the range of the abscissa.


%
%


Although we have found $D_{\textrm{c}}$ and $D_{\textrm{h}}$ from the largest clusters,
analogous results are obtained for the dimensions of the whole domain morphology at scales larger than $\sqrt{\lambda_{\textrm{d,h}}t}$ [Fig. S2, Suppl. Mat.].


\begin{figure}[!b]
\includegraphics[width=0.49\columnwidth]{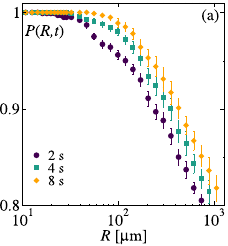} 
\includegraphics[width=0.49\columnwidth]{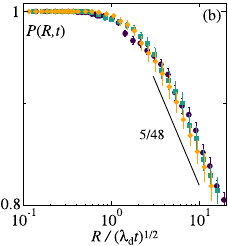}
\caption{\label{fig_probradius}
Probability of finding a cluster connecting the region around the origin to the circle of radius $R$
as a function of (a) $R$ and (b) $R/\sqrt{\lambda_{\textrm{d}}t}$ in semi and double logarithmic scale, respectively;
$\lambda_{\textrm{d}} = \unit[767]{\mu m^{2}s^{-1}}$.
Solid line indicates the correspondent one-arm exponent for percolation.
Uncertainties are one standard deviation of the mean.}
\end{figure}


\textit{Crossing events}.--- 
A fundamental observable in percolation theory is the probability $P(R)$ that 
there is a cluster connecting a region in the vicinity of the origin to the circle of radius $R$~\cite{Smirnov01_exponents}.
For critical site percolation on the triangular grid, 
the Lawler-Schramm-Werner (LSW) theorem~\cite{Lawler02_paper4} states that 
$P(R) = (R/R_0)^{-\zeta + o(1)}$ 
for $R \gg R_0$, where $\zeta = 5/48$ is the one-arm exponent, and $R_0$ is a microscopic length.
The law for $P(R)$ should hold for other lattice geometries on the basis of the universality hypothesis.
%
%
For the ordering problem, 
we can regard the domain morphology of a curvature-driven process at $t \gg t_{\textrm{p}}$ 
as equivalent to that of CP with enlarging lattice spacing $R_0(t) \sim \sqrt{t}$~\cite{Arenzon07, Corberi17}.
Thus, we expect $P(R,t) \simeq h(R/\sqrt{\lambda_\textrm{d}t})$ with $h(y) \sim y^{-\zeta}$ for $y \gg 1$.
If percolation events are not distinguished by the phase of the spanning cluster, then $h(y) = 1$ for $y \ll 1$.


To analyze the proposed form, we locate the origin at the middle point of the images.
From there, 
we draw a circle 
$C_{\textrm{R}}$ 
of radius $R \in [a, (L_{\textrm{x}}/2) - a]$ 
by solving circle's equation with increments of $\delta = 10^{-4}$ (smaller increments do not affect the results). 
Pixels touched by $C_{\textrm{R}}$ define a frontier region.
We are interested in the probability of finding 
a connected path between the region of area $a^2$ around the origin (i.e., the central pixel) and the frontier region.
To achieve that, we distribute the ensemble of images at $t$ in 10 independent groups of $10^2$ images each.
For each group, we measured the empirical probability for the percolation event of interest.
After taking the average among the set of empirical probabilities, we obtain $P(R,t)$.


Figure~\ref{fig_probradius}(a) shows $P(R,t)$ as a function of $R$.
The expected initial regime in which $P = 1$ at short radii occurs up to a time-dependent scale $\sim \sqrt{t}$.
Afterwards, the curves have an extremely slow decay, monotonically decreasing from 1 to 0.8 while $R$ varies two decades.
A good data collapse is obtained when the data are plotted versus $y\equiv R/\sqrt{\lambda_{\textrm{d}}t}$, Fig.~\ref{fig_probradius}(b).
Remarkably, the slow decay is well described by the algebraic law, $y^{-5/48}$, inherited from the LSW theorem.
%


Results hitherto indicate that the scaling relation $D_{\textrm{c}} + \zeta = 2$~\cite{Stauffer_book} for percolation models is also satisfied in the nonequilibrium system.
This allows us to identify the one-arm exponent in the time dependence for the mass of the largest cluster as, 
according to the scaling function in Fig.~\ref{fig_massscaling}(a),
$M_{\textrm{c}}(L,t)a^2 \sim  L^{D_{\textrm{c}}}t^{\zeta/2}$ for $L \gg \sqrt{\lambda_{\textrm{d}}t}$.


\textit{SLE diffusivity}.--- 
How tortuous are the fractal contours of domains, such as those shown in Fig.~\ref{fig_cluster}(c) and Fig.~\ref{fig_cluster}(d)?
An answer can be given in terms of winding angle measurements.
The winding angle $\theta$, $\theta \in \mathbb{R}$, 
of a path in the plane is the net rotation performed by the tangent to the path when one continuously moves along it.
For conformally invariant curves described by SLE,
it is expected~\cite{Duplantier88, Wieland03} (see also Refs.~\cite{Schramm00, Beffara08_Hausdorff}) that $\theta$ has null average 
and variance given by:

    \begin{equation}
    \langle \theta^2(w) \rangle = c + \bigg(\frac{4\kappa}{8+\kappa}\bigg)\ln{w},
    \label{eq_wangle}
    \end{equation}
where $w$ is a measure of the curvilinear distance along the path and $c$ is a non-universal constant. 
For the ordering problem, interest is on its dynamical counterpart, $\langle \theta^2(w,t) \rangle$ \cite{Corberi17, Blanchard17}.
The SLE diffusivity, $\kappa$ \cite{Schramm00}, shall not be confused with the local curvature (same symbol) appearing in the interfacial equation of motion noted in Ref.~\cite{Almeida23}.
%
%


To test Eq.~\eqref{eq_wangle}, we proceed as follows.
For each sample at $t$, we extract the hull of the largest domain that does not touch a border of the image.
In the discrete approximation for the experimental data, the interface of a hull can be defined as the set of edges, of each pixel, facing the exterior of the hull.
The length of an interface, or equivalently the perimeter $p$ of its parent hull, is the number $N$ of such edges times $a$.
Interfaces are oriented according to a phase rule~\cite{Tartaglia18}:
To their right side there is a domain of phase $s = +1$.
This implies that the $i^{\rm th}$ edge of an interface receives the orientation variable $\chi_i$, $i \in [1,N]$.
On the square mesh, there are only four possible outcomes for $\chi$: $\leftarrow$, $\uparrow$, $\rightarrow$, $\downarrow$. 
The turning angle $\alpha_i$ is the angle between $\chi_i$ and $\chi_{i+1}$ measured clockwise in radians, $i < N$;
$\alpha_N$ is the angle between $\chi_N$ and $\chi_{1}$ if the $N^{\rm th}$ edge is connected to the $1^{\rm st}$ edge. 
For a connected path of size $\ell = na$, $n \in \mathbb{N}^{+}$ such that $n \leq N$, $\theta(\ell) = \sum_{j=1}^{n} \alpha_j$. 
Since the interfaces we consider are closed loops having no crossing points, $\theta = \pm 2\pi$ for $\ell = Na = p$.


\begin{figure}[!t]
\includegraphics[width=0.49\columnwidth]{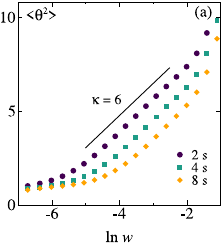} 
\includegraphics[width=0.49\columnwidth]{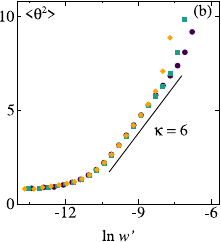}
\caption{\label{fig_wangle} (a) Average squared winding angle of loops as function of the logarithm of the normalized curvilinear distance.
(b) Evidence for dynamical scaling.
Solid lines show the slope expected from Schramm-Loewner evolution process with $\kappa = 6$. 
Plots have the same keys.
Uncertainties are one standard deviation of the mean.
}\end{figure}


For each loop, we average $\theta(\ell,t)$ and $\theta^2$ over all possible paths of length $\ell$.
To further perform averages over the ensemble of loops at $t$,
we use a normalized curvilinear distance, here $w = \ell/p$, since the curvature and length of loops fluctuate from sample to sample.
Experimental outcomes are shown in Fig.~\ref{fig_wangle}(a).
The $\langle \theta^2(w,t) \rangle$ curves comprise three regimes along the increasing scale of $\ln{w}$.
At short scales, $\langle \theta^2 \rangle \approx 1$ with a subtle lift because interfaces tend to be locally flat.
At the intermediate regime, $\langle \theta^2 \rangle$ grows linearly with $\ln{w}$ in the same way that the family of SLE curves with parameter $\kappa = 6$ does, Eq.~\eqref{eq_wangle}; $\langle \theta \rangle \approx 0$ in the same interval (not shown).
The intermediate part finishes once the curvature of the path induces $\langle \theta^2 \rangle$ to rapidly increase towards its boundary value $(\pm 2\pi)^2$.
As time elapses, the first and the third regimes stretch at the expense of the intermediate region.
The former is due the growth of correlations up to the order of the coarsening length, 
which act to morph winding boundaries into flatter segments. 
The latter occurs because loops are shrinking.


According to the dynamical scaling hypothesis, the data in Fig.~\ref{fig_wangle}(a) shall fall on a master curve when plotted as function of $\ell/\sqrt{\lambda_{\textrm{h}}t}$. 
For a fractal hull of linear size $L'$, $p/\sqrt{\lambda_{\textrm{h}}t} \sim (L'/\sqrt{\lambda_{\textrm{h}}t})^{D_{\textrm{h}}}$.
Replacing $L'$ (a variable that depends on each loop) by the pixel size $a$, we can define
$w' = (\ell/p) \times (a / \sqrt{\lambda_{\textrm{h}}t})^{D_{\textrm{h}}}$ as a dimensionless variable, 
$[w'] = [\ell]/[\sqrt{\lambda_{\textrm{h}}t}]$,
that allows us to map the data in Fig.~\ref{fig_wangle}(a) into the scaling function shown in Fig.~\ref{fig_wangle}(b); 
we recall that $D_{\textrm{h}} = 7/4$.
The experimental data clearly respect the dynamic scaling hypothesis.
The collapse into a master curve takes place up to the non-logarithmic regime.
The intermediate regime preserves the logarithmic growth, Eq.~\eqref{eq_wangle} with $w \mapsto w'$.
By reading $d\langle \theta^2 \rangle/d \ln{w'}$ of the data at $t = \unit[2.0]{s}$ and averaging over $-10 < \ln{w'} < -7.5$, we obtain $4\kappa/(8+\kappa) = 1.74 \pm 0.04$.
This gives $\kappa = 6.2 \pm 0.4$, in good agreement with $\kappa = 6$ for hulls in the percolation universality class~\cite{Smirnov01}.
At certain scales, hulls in the liquid crystalline material behave as conformal invariant curves.
%


\textit{Conclusions}.---
We have studied the two-dimensional domain morphology of twisted nematic phases during their ordering kinetics~\cite{Almeida23} by focusing on the geometry of clusters. 
The largest nematic domains and their hulls have fractal dimensions in agreement with those of the corresponding figures in critical percolation.
Their dimensions also describe the whole domain morphology at scales larger than the coarsening length.
The Lawler-Schramm-Werner theorem~\cite{Lawler02_paper4}, originally conceived for site percolation on the triangular grid, can elegantly predict
the decay of a crossing probability from the center to the boundary of disks in the experimental data
(provided that a microscopic length in the original formulation is interpreted as the typical domain size in the nonequilibrium problem).
The observation of a logarithmic growth for the squared winding angle of hulls, Eq.~\eqref{eq_wangle}, 
indicates that the contours of twisted nematic regions behave as conformally invariant curves at such scales.
This logarithmic growth is compatible with that for the family of SLE curves with diffusivity $\kappa = 6$.
These results show an experimental realization of critical clusters in the percolation universality class. 
They demonstrate the empirical relevance of theoretical and mathematical results which are remarkable
recent achievements in Statistical Mechanics and Probability theory.
A new path for experimentally accessing
the rich predictions of conformally invariant and SLE$_6$ curves is open.

\begin{acknowledgments}
%
Authors were partially funded by the Brazilian National Council for Scientific and Technological Development – CNPq, Coordenação de
Aperfeiçoamento de Pessoal de Nível Superior – CAPES
(Finance Code 001), and the Fundação de Amparo à
Pesquisa do Estado do Rio Grande do Sul – FAPERGS,
Grant No. 23/2551-0000154-3. R. A. L. A. was also funded
by the Japan Society for the Promotion of Science – JSPS,
Grant No. JP16J06923. J. J. A. acknowledges CNPq,
Grants No. 316628/2021-2, No. 402487/2023-0, No. and
443517/2023-1.
Data and codes are available upon reasonable request.
\end{acknowledgments}


\begin{thebibliography}{41}%
\makeatletter
\providecommand \@ifxundefined [1]{%
 \@ifx{#1\undefined}
}%
\providecommand \@ifnum [1]{%
 \ifnum #1\expandafter \@firstoftwo
 \else \expandafter \@secondoftwo
 \fi
}%
\providecommand \@ifx [1]{%
 \ifx #1\expandafter \@firstoftwo
 \else \expandafter \@secondoftwo
 \fi
}%
\providecommand \natexlab [1]{#1}%
\providecommand \enquote  [1]{``#1''}%
\providecommand \bibnamefont  [1]{#1}%
\providecommand \bibfnamefont [1]{#1}%
\providecommand \citenamefont [1]{#1}%
\providecommand \href@noop [0]{\@secondoftwo}%
\providecommand \href [0]{\begingroup \@sanitize@url \@href}%
\providecommand \@href[1]{\@@startlink{#1}\@@href}%
\providecommand \@@href[1]{\endgroup#1\@@endlink}%
\providecommand \@sanitize@url [0]{\catcode `\\12\catcode `\$12\catcode
  `\&12\catcode `\#12\catcode `\^12\catcode `\_12\catcode `\%12\relax}%
\providecommand \@@startlink[1]{}%
\providecommand \@@endlink[0]{}%
\providecommand \url  [0]{\begingroup\@sanitize@url \@url }%
\providecommand \@url [1]{\endgroup\@href {#1}{\urlprefix }}%
\providecommand \urlprefix  [0]{URL }%
\providecommand \Eprint [0]{\href }%
\providecommand \doibase [0]{https://doi.org/}%
\providecommand \selectlanguage [0]{\@gobble}%
\providecommand \bibinfo  [0]{\@secondoftwo}%
\providecommand \bibfield  [0]{\@secondoftwo}%
\providecommand \translation [1]{[#1]}%
\providecommand \BibitemOpen [0]{}%
\providecommand \bibitemStop [0]{}%
\providecommand \bibitemNoStop [0]{.\EOS\space}%
\providecommand \EOS [0]{\spacefactor3000\relax}%
\providecommand \BibitemShut  [1]{\csname bibitem#1\endcsname}%
\let\auto@bib@innerbib\@empty
\bibitem [{\citenamefont {Bray}(2002)}]{Bray02}%
  \BibitemOpen
  \bibfield  {author} {\bibinfo {author} {\bibfnamefont {A.~J.}\ \bibnamefont
  {Bray}},\ }\bibfield  {title} {\bibinfo {title} {{Theory of phase-ordering
  kinetics}},\ }\href {https://doi.org/10.1080/00018730110117433} {\bibfield
  {journal} {\bibinfo  {journal} {Adv. Phys.}\ }\textbf {\bibinfo {volume}
  {51}},\ \bibinfo {pages} {481} (\bibinfo {year} {2002})}\BibitemShut
  {NoStop}%
\bibitem [{\citenamefont {Cugliandolo}(2015)}]{Cugliandolo15}%
  \BibitemOpen
  \bibfield  {author} {\bibinfo {author} {\bibfnamefont {L.~F.}\ \bibnamefont
  {Cugliandolo}},\ }\bibfield  {title} {\bibinfo {title} {{Coarsening
  phenomena}},\ }\href
  {https://doi.org/https://doi.org/10.1016/j.crhy.2015.02.005} {\bibfield
  {journal} {\bibinfo  {journal} {Comptes Rendus Physique}\ }\textbf {\bibinfo
  {volume} {16}},\ \bibinfo {pages} {257} (\bibinfo {year} {2015})}\BibitemShut
  {NoStop}%
\bibitem [{\citenamefont {Sicilia}\ \emph
  {et~al.}(2008{\natexlab{a}})\citenamefont {Sicilia} \emph
  {et~al.}}]{Sicilia08}%
  \BibitemOpen
  \bibfield  {author} {\bibinfo {author} {\bibfnamefont {A.}~\bibnamefont
  {Sicilia}} \emph {et~al.},\ }\bibfield  {title} {\bibinfo {title}
  {{Experimental Test of Curvature-Driven Dynamics in the Phase Ordering of a
  Two Dimensional Liquid Crystal}},\ }\href
  {https://doi.org/10.1103/PhysRevLett.101.197801} {\bibfield  {journal}
  {\bibinfo  {journal} {Phys. Rev. Lett.}\ }\textbf {\bibinfo {volume} {101}},\
  \bibinfo {pages} {197801} (\bibinfo {year} {2008}{\natexlab{a}})}\BibitemShut
  {NoStop}%
\bibitem [{\citenamefont {Almeida}(2023)}]{Almeida23}%
  \BibitemOpen
  \bibfield  {author} {\bibinfo {author} {\bibfnamefont {R.~A.~L.}\
  \bibnamefont {Almeida}},\ }\bibfield  {title} {\bibinfo {title} {{Critical
  Percolation in the Ordering Kinetics of Twisted Nematic Phases}},\ }\href
  {https://doi.org/10.1103/PhysRevLett.131.268101} {\bibfield  {journal}
  {\bibinfo  {journal} {Phys. Rev. Lett.}\ }\textbf {\bibinfo {volume} {131}},\
  \bibinfo {pages} {268101} (\bibinfo {year} {2023})}\BibitemShut {NoStop}%
\bibitem [{\citenamefont {McNally}\ \emph {et~al.}(2017)\citenamefont {McNally}
  \emph {et~al.}}]{Mcnally17}%
  \BibitemOpen
  \bibfield  {author} {\bibinfo {author} {\bibfnamefont {L.}~\bibnamefont
  {McNally}} \emph {et~al.},\ }\bibfield  {title} {\bibinfo {title} {{Killing
  by Type VI secretion drives genetic phase separation and correlates with
  increased cooperation}},\ }\href {https://doi.org/10.1038/ncomms14371}
  {\bibfield  {journal} {\bibinfo  {journal} {Nature communications}\ }\textbf
  {\bibinfo {volume} {8}},\ \bibinfo {pages} {14371} (\bibinfo {year}
  {2017})}\BibitemShut {NoStop}%
\bibitem [{\citenamefont {Andersen}\ \emph {et~al.}(2024)\citenamefont
  {Andersen} \emph {et~al.}}]{Andersen24}%
  \BibitemOpen
  \bibfield  {author} {\bibinfo {author} {\bibfnamefont {B.~H.}\ \bibnamefont
  {Andersen}} \emph {et~al.},\ }\bibfield  {title} {\bibinfo {title} {Evidence
  of robust, universal conformal invariance in living biological matter},\
  }\href@noop {} {\bibfield  {journal} {\bibinfo  {journal} {arXiv preprint
  arXiv:2403.08466}\ } (\bibinfo {year} {2024})}\BibitemShut {NoStop}%
\bibitem [{\citenamefont {Arenzon}\ \emph {et~al.}(2007)\citenamefont
  {Arenzon}, \citenamefont {Bray}, \citenamefont {Cugliandolo},\ and\
  \citenamefont {Sicilia}}]{Arenzon07}%
  \BibitemOpen
  \bibfield  {author} {\bibinfo {author} {\bibfnamefont {J.~J.}\ \bibnamefont
  {Arenzon}}, \bibinfo {author} {\bibfnamefont {A.~J.}\ \bibnamefont {Bray}},
  \bibinfo {author} {\bibfnamefont {L.~F.}\ \bibnamefont {Cugliandolo}},\ and\
  \bibinfo {author} {\bibfnamefont {A.}~\bibnamefont {Sicilia}},\ }\bibfield
  {title} {\bibinfo {title} {{Exact Results for Curvature-Driven Coarsening in
  Two Dimensions}},\ }\href {https://doi.org/10.1103/PhysRevLett.98.145701}
  {\bibfield  {journal} {\bibinfo  {journal} {Phys. Rev. Lett.}\ }\textbf
  {\bibinfo {volume} {98}},\ \bibinfo {pages} {145701} (\bibinfo {year}
  {2007})}\BibitemShut {NoStop}%
\bibitem [{\citenamefont {Hofmann}\ \emph {et~al.}(2014)\citenamefont
  {Hofmann}, \citenamefont {Natu},\ and\ \citenamefont
  {Das~Sarma}}]{Hofmann14}%
  \BibitemOpen
  \bibfield  {author} {\bibinfo {author} {\bibfnamefont {J.}~\bibnamefont
  {Hofmann}}, \bibinfo {author} {\bibfnamefont {S.~S.}\ \bibnamefont {Natu}},\
  and\ \bibinfo {author} {\bibfnamefont {S.}~\bibnamefont {Das~Sarma}},\
  }\bibfield  {title} {\bibinfo {title} {{Coarsening Dynamics of Binary Bose
  Condensates}},\ }\href {https://doi.org/10.1103/PhysRevLett.113.095702}
  {\bibfield  {journal} {\bibinfo  {journal} {Phys. Rev. Lett.}\ }\textbf
  {\bibinfo {volume} {113}},\ \bibinfo {pages} {095702} (\bibinfo {year}
  {2014})}\BibitemShut {NoStop}%
\bibitem [{\citenamefont {Dornic}\ \emph {et~al.}(2001)\citenamefont {Dornic},
  \citenamefont {Chat\'e}, \citenamefont {Chave},\ and\ \citenamefont
  {Hinrichsen}}]{Dornic01}%
  \BibitemOpen
  \bibfield  {author} {\bibinfo {author} {\bibfnamefont {I.}~\bibnamefont
  {Dornic}}, \bibinfo {author} {\bibfnamefont {H.}~\bibnamefont {Chat\'e}},
  \bibinfo {author} {\bibfnamefont {J.}~\bibnamefont {Chave}},\ and\ \bibinfo
  {author} {\bibfnamefont {H.}~\bibnamefont {Hinrichsen}},\ }\bibfield  {title}
  {\bibinfo {title} {{Critical Coarsening without Surface Tension: The
  Universality Class of the Voter Model}},\ }\href
  {https://doi.org/10.1103/PhysRevLett.87.045701} {\bibfield  {journal}
  {\bibinfo  {journal} {Phys. Rev. Lett.}\ }\textbf {\bibinfo {volume} {87}},\
  \bibinfo {pages} {045701} (\bibinfo {year} {2001})}\BibitemShut {NoStop}%
\bibitem [{\citenamefont {Latoski}\ \emph {et~al.}(2024)\citenamefont
  {Latoski}, \citenamefont {Dantas},\ and\ \citenamefont
  {Arenzon}}]{Latoski24}%
  \BibitemOpen
  \bibfield  {author} {\bibinfo {author} {\bibfnamefont {L.~C.~F.}\
  \bibnamefont {Latoski}}, \bibinfo {author} {\bibfnamefont {W.~G.}\
  \bibnamefont {Dantas}},\ and\ \bibinfo {author} {\bibfnamefont {J.~J.}\
  \bibnamefont {Arenzon}},\ }\bibfield  {title} {\bibinfo {title} {{Opinion
  inertia and coarsening in the persistent voter model}},\ }\href
  {https://doi.org/10.1103/PhysRevE.109.054115} {\bibfield  {journal} {\bibinfo
   {journal} {Phys. Rev. E}\ }\textbf {\bibinfo {volume} {109}},\ \bibinfo
  {pages} {054115} (\bibinfo {year} {2024})}\BibitemShut {NoStop}%
\bibitem [{\citenamefont {Barros}\ \emph {et~al.}(2009)\citenamefont {Barros},
  \citenamefont {Krapivsky},\ and\ \citenamefont {Redner}}]{Barros09}%
  \BibitemOpen
  \bibfield  {author} {\bibinfo {author} {\bibfnamefont {K.}~\bibnamefont
  {Barros}}, \bibinfo {author} {\bibfnamefont {P.~L.}\ \bibnamefont
  {Krapivsky}},\ and\ \bibinfo {author} {\bibfnamefont {S.}~\bibnamefont
  {Redner}},\ }\bibfield  {title} {\bibinfo {title} {{Freezing into stripe
  states in two-dimensional ferromagnets and crossing probabilities in critical
  percolation}},\ }\href {https://doi.org/10.1103/PhysRevE.80.040101}
  {\bibfield  {journal} {\bibinfo  {journal} {Phys. Rev. E}\ }\textbf {\bibinfo
  {volume} {80}},\ \bibinfo {pages} {040101(R)} (\bibinfo {year}
  {2009})}\BibitemShut {NoStop}%
\bibitem [{\citenamefont {Olejarz}\ \emph {et~al.}(2012)\citenamefont
  {Olejarz}, \citenamefont {Krapivsky},\ and\ \citenamefont
  {Redner}}]{Olejarz12}%
  \BibitemOpen
  \bibfield  {author} {\bibinfo {author} {\bibfnamefont {J.}~\bibnamefont
  {Olejarz}}, \bibinfo {author} {\bibfnamefont {P.~L.}\ \bibnamefont
  {Krapivsky}},\ and\ \bibinfo {author} {\bibfnamefont {S.}~\bibnamefont
  {Redner}},\ }\bibfield  {title} {\bibinfo {title} {{Fate of 2D Kinetic
  Ferromagnets and Critical Percolation Crossing Probabilities}},\ }\href
  {https://doi.org/10.1103/PhysRevLett.109.195702} {\bibfield  {journal}
  {\bibinfo  {journal} {Phys. Rev. Lett.}\ }\textbf {\bibinfo {volume} {109}},\
  \bibinfo {pages} {195702} (\bibinfo {year} {2012})}\BibitemShut {NoStop}%
\bibitem [{\citenamefont {Stauffer}\ and\ \citenamefont
  {Aharony}(1994)}]{Stauffer_book}%
  \BibitemOpen
  \bibfield  {author} {\bibinfo {author} {\bibfnamefont {D.}~\bibnamefont
  {Stauffer}}\ and\ \bibinfo {author} {\bibfnamefont {A.}~\bibnamefont
  {Aharony}},\ }\href@noop {} {\emph {\bibinfo {title} {Introduction to
  percolation theory}}}\ (\bibinfo  {publisher} {Taylor \& Francis},\ \bibinfo
  {address} {London},\ \bibinfo {year} {1994})\BibitemShut {NoStop}%
\bibitem [{\citenamefont {Araújo}\ \emph {et~al.}(2014)\citenamefont
  {Araújo}, \citenamefont {Grassberger}, \citenamefont {Kahng}, \citenamefont
  {Schrenk},\ and\ \citenamefont {Ziff}}]{Araujo14}%
  \BibitemOpen
  \bibfield  {author} {\bibinfo {author} {\bibfnamefont {N.}~\bibnamefont
  {Araújo}}, \bibinfo {author} {\bibfnamefont {P.}~\bibnamefont
  {Grassberger}}, \bibinfo {author} {\bibfnamefont {B.}~\bibnamefont {Kahng}},
  \bibinfo {author} {\bibfnamefont {K.}~\bibnamefont {Schrenk}},\ and\ \bibinfo
  {author} {\bibfnamefont {R.}~\bibnamefont {Ziff}},\ }\bibfield  {title}
  {\bibinfo {title} {Recent advances and open challenges in percolation},\
  }\href {https://doi.org/10.1140/epjst/e2014-02266-y} {\bibfield  {journal}
  {\bibinfo  {journal} {Eur. Phys. J. Spec. Top.}\ }\textbf {\bibinfo {volume}
  {223}},\ \bibinfo {pages} {2307} (\bibinfo {year} {2014})}\BibitemShut
  {NoStop}%
\bibitem [{\citenamefont {Saberi}(2015)}]{Saberi15}%
  \BibitemOpen
  \bibfield  {author} {\bibinfo {author} {\bibfnamefont {A.~A.}\ \bibnamefont
  {Saberi}},\ }\bibfield  {title} {\bibinfo {title} {{Recent advances in
  percolation theory and its applications}},\ }\href
  {https://doi.org/https://doi.org/10.1016/j.physrep.2015.03.003} {\bibfield
  {journal} {\bibinfo  {journal} {Phys. Rep.}\ }\textbf {\bibinfo {volume}
  {578}},\ \bibinfo {pages} {1} (\bibinfo {year} {2015})}\BibitemShut {NoStop}%
\bibitem [{\citenamefont {Blanchard}\ \emph {et~al.}(2014)\citenamefont
  {Blanchard}, \citenamefont {Corberi}, \citenamefont {Cugliandolo},\ and\
  \citenamefont {Picco}}]{Blanchard14}%
  \BibitemOpen
  \bibfield  {author} {\bibinfo {author} {\bibfnamefont {T.}~\bibnamefont
  {Blanchard}}, \bibinfo {author} {\bibfnamefont {F.}~\bibnamefont {Corberi}},
  \bibinfo {author} {\bibfnamefont {L.~F.}\ \bibnamefont {Cugliandolo}},\ and\
  \bibinfo {author} {\bibfnamefont {M.}~\bibnamefont {Picco}},\ }\bibfield
  {title} {\bibinfo {title} {{How soon after a zero-temperature quench is the
  fate of the Ising model sealed?}},\ }\href
  {https://doi.org/10.1209/0295-5075/106/66001} {\bibfield  {journal} {\bibinfo
   {journal} {Europhys. Lett.}\ }\textbf {\bibinfo {volume} {106}},\ \bibinfo
  {pages} {66001} (\bibinfo {year} {2014})}\BibitemShut {NoStop}%
\bibitem [{\citenamefont {de~Azevedo-Lopes}\ \emph {et~al.}(2022)\citenamefont
  {de~Azevedo-Lopes}, \citenamefont {Almeida}, \citenamefont {de~Oliveira},\
  and\ \citenamefont {Arenzon}}]{Azevedo22}%
  \BibitemOpen
  \bibfield  {author} {\bibinfo {author} {\bibfnamefont {A.}~\bibnamefont
  {de~Azevedo-Lopes}}, \bibinfo {author} {\bibfnamefont {R.~A.~L.}\
  \bibnamefont {Almeida}}, \bibinfo {author} {\bibfnamefont {P.~M.~C.}\
  \bibnamefont {de~Oliveira}},\ and\ \bibinfo {author} {\bibfnamefont {J.~J.}\
  \bibnamefont {Arenzon}},\ }\bibfield  {title} {\bibinfo {title}
  {{Energy-lowering and constant-energy spin flips: Emergence of the
  percolating cluster in the kinetic Ising model}},\ }\href
  {https://doi.org/10.1103/PhysRevE.106.044105} {\bibfield  {journal} {\bibinfo
   {journal} {Phys. Rev. E}\ }\textbf {\bibinfo {volume} {106}},\ \bibinfo
  {pages} {044105} (\bibinfo {year} {2022})}\BibitemShut {NoStop}%
\bibitem [{\citenamefont {Sicilia}\ \emph {et~al.}(2007)\citenamefont
  {Sicilia}, \citenamefont {Arenzon}, \citenamefont {Bray},\ and\ \citenamefont
  {Cugliandolo}}]{Sicilia07}%
  \BibitemOpen
  \bibfield  {author} {\bibinfo {author} {\bibfnamefont {A.}~\bibnamefont
  {Sicilia}}, \bibinfo {author} {\bibfnamefont {J.~J.}\ \bibnamefont
  {Arenzon}}, \bibinfo {author} {\bibfnamefont {A.~J.}\ \bibnamefont {Bray}},\
  and\ \bibinfo {author} {\bibfnamefont {L.~F.}\ \bibnamefont {Cugliandolo}},\
  }\bibfield  {title} {\bibinfo {title} {{Domain growth morphology in
  curvature-driven two-dimensional coarsening}},\ }\href
  {https://doi.org/10.1103/PhysRevE.76.061116} {\bibfield  {journal} {\bibinfo
  {journal} {Phys. Rev. E}\ }\textbf {\bibinfo {volume} {76}},\ \bibinfo
  {pages} {061116} (\bibinfo {year} {2007})}\BibitemShut {NoStop}%
\bibitem [{\citenamefont {Sicilia}\ \emph {et~al.}(2009)\citenamefont
  {Sicilia}, \citenamefont {Sarrazin}, \citenamefont {Arenzon}, \citenamefont
  {Bray},\ and\ \citenamefont {Cugliandolo}}]{Sicilia09}%
  \BibitemOpen
  \bibfield  {author} {\bibinfo {author} {\bibfnamefont {A.}~\bibnamefont
  {Sicilia}}, \bibinfo {author} {\bibfnamefont {Y.}~\bibnamefont {Sarrazin}},
  \bibinfo {author} {\bibfnamefont {J.~J.}\ \bibnamefont {Arenzon}}, \bibinfo
  {author} {\bibfnamefont {A.~J.}\ \bibnamefont {Bray}},\ and\ \bibinfo
  {author} {\bibfnamefont {L.~F.}\ \bibnamefont {Cugliandolo}},\ }\bibfield
  {title} {\bibinfo {title} {Geometry of phase separation},\ }\href
  {https://doi.org/10.1103/PhysRevE.80.031121} {\bibfield  {journal} {\bibinfo
  {journal} {Phys. Rev. E}\ }\textbf {\bibinfo {volume} {80}},\ \bibinfo
  {pages} {031121} (\bibinfo {year} {2009})}\BibitemShut {NoStop}%
\bibitem [{\citenamefont {Tartaglia}\ \emph {et~al.}(2018)\citenamefont
  {Tartaglia}, \citenamefont {Cugliandolo},\ and\ \citenamefont
  {Picco}}]{Tartaglia18}%
  \BibitemOpen
  \bibfield  {author} {\bibinfo {author} {\bibfnamefont {A.}~\bibnamefont
  {Tartaglia}}, \bibinfo {author} {\bibfnamefont {L.~F.}\ \bibnamefont
  {Cugliandolo}},\ and\ \bibinfo {author} {\bibfnamefont {M.}~\bibnamefont
  {Picco}},\ }\bibfield  {title} {\bibinfo {title} {{Coarsening and percolation
  in the kinetic 2d Ising model with spin exchange updates and the voter
  model}},\ }\href {https://doi.org/10.1088/1742-5468/aad366} {\bibfield
  {journal} {\bibinfo  {journal} {J. Stat. Mech.}\ }\textbf {\bibinfo {volume}
  {2018}},\ \bibinfo {pages} {083202} (\bibinfo {year} {2018})}\BibitemShut
  {NoStop}%
\bibitem [{\citenamefont {Sicilia}\ \emph
  {et~al.}(2008{\natexlab{b}})\citenamefont {Sicilia}, \citenamefont {Arenzon},
  \citenamefont {Bray},\ and\ \citenamefont
  {Cugliandolo}}]{Sicilia08_disorder}%
  \BibitemOpen
  \bibfield  {author} {\bibinfo {author} {\bibfnamefont {A.}~\bibnamefont
  {Sicilia}}, \bibinfo {author} {\bibfnamefont {J.~J.}\ \bibnamefont
  {Arenzon}}, \bibinfo {author} {\bibfnamefont {A.~J.}\ \bibnamefont {Bray}},\
  and\ \bibinfo {author} {\bibfnamefont {L.~F.}\ \bibnamefont {Cugliandolo}},\
  }\bibfield  {title} {\bibinfo {title} {{Geometric properties of
  two-dimensional coarsening with weak disorder}},\ }\href
  {https://doi.org/10.1209/0295-5075/82/10001} {\bibfield  {journal} {\bibinfo
  {journal} {Europhys. Lett.}\ }\textbf {\bibinfo {volume} {82}},\ \bibinfo
  {pages} {10001} (\bibinfo {year} {2008}{\natexlab{b}})}\BibitemShut {NoStop}%
\bibitem [{\citenamefont {Corberi}\ \emph {et~al.}(2017)\citenamefont
  {Corberi}, \citenamefont {Cugliandolo}, \citenamefont {Insalata},\ and\
  \citenamefont {Picco}}]{Corberi17}%
  \BibitemOpen
  \bibfield  {author} {\bibinfo {author} {\bibfnamefont {F.}~\bibnamefont
  {Corberi}}, \bibinfo {author} {\bibfnamefont {L.~F.}\ \bibnamefont
  {Cugliandolo}}, \bibinfo {author} {\bibfnamefont {F.}~\bibnamefont
  {Insalata}},\ and\ \bibinfo {author} {\bibfnamefont {M.}~\bibnamefont
  {Picco}},\ }\bibfield  {title} {\bibinfo {title} {{Coarsening and percolation
  in a disordered ferromagnet}},\ }\href
  {https://doi.org/10.1103/PhysRevE.95.022101} {\bibfield  {journal} {\bibinfo
  {journal} {Phys. Rev. E}\ }\textbf {\bibinfo {volume} {95}},\ \bibinfo
  {pages} {022101} (\bibinfo {year} {2017})}\BibitemShut {NoStop}%
\bibitem [{\citenamefont {Ricateau}\ \emph {et~al.}(2018)\citenamefont
  {Ricateau}, \citenamefont {Cugliandolo},\ and\ \citenamefont
  {Picco}}]{Ricateau18}%
  \BibitemOpen
  \bibfield  {author} {\bibinfo {author} {\bibfnamefont {H.}~\bibnamefont
  {Ricateau}}, \bibinfo {author} {\bibfnamefont {L.~F.}\ \bibnamefont
  {Cugliandolo}},\ and\ \bibinfo {author} {\bibfnamefont {M.}~\bibnamefont
  {Picco}},\ }\bibfield  {title} {\bibinfo {title} {{Critical percolation in
  the slow cooling of the bi-dimensional ferromagnetic Ising model}},\ }\href
  {https://doi.org/10.1088/1742-5468/aa9bb4} {\bibfield  {journal} {\bibinfo
  {journal} {J. Stat. Mech.}\ }\textbf {\bibinfo {volume} {2018}},\ \bibinfo
  {pages} {013201} (\bibinfo {year} {2018})}\BibitemShut {NoStop}%
\bibitem [{\citenamefont {Agrawal}\ \emph {et~al.}(2022)\citenamefont
  {Agrawal}, \citenamefont {Corberi}, \citenamefont {Insalata},\ and\
  \citenamefont {Puri}}]{Agrawal22}%
  \BibitemOpen
  \bibfield  {author} {\bibinfo {author} {\bibfnamefont {R.}~\bibnamefont
  {Agrawal}}, \bibinfo {author} {\bibfnamefont {F.}~\bibnamefont {Corberi}},
  \bibinfo {author} {\bibfnamefont {F.}~\bibnamefont {Insalata}},\ and\
  \bibinfo {author} {\bibfnamefont {S.}~\bibnamefont {Puri}},\ }\bibfield
  {title} {\bibinfo {title} {{Asymptotic states of Ising ferromagnets with
  long-range interactions}},\ }\href
  {https://doi.org/10.1103/PhysRevE.105.034131} {\bibfield  {journal} {\bibinfo
   {journal} {Phys. Rev. E}\ }\textbf {\bibinfo {volume} {105}},\ \bibinfo
  {pages} {034131} (\bibinfo {year} {2022})}\BibitemShut {NoStop}%
\bibitem [{\citenamefont {Agrawal}\ \emph {et~al.}(2023)\citenamefont
  {Agrawal}, \citenamefont {Cugliandolo}, \citenamefont {Faoro}, \citenamefont
  {Ioffe},\ and\ \citenamefont {Picco}}]{Agrawal23}%
  \BibitemOpen
  \bibfield  {author} {\bibinfo {author} {\bibfnamefont {R.}~\bibnamefont
  {Agrawal}}, \bibinfo {author} {\bibfnamefont {L.~F.}\ \bibnamefont
  {Cugliandolo}}, \bibinfo {author} {\bibfnamefont {L.}~\bibnamefont {Faoro}},
  \bibinfo {author} {\bibfnamefont {L.~B.}\ \bibnamefont {Ioffe}},\ and\
  \bibinfo {author} {\bibfnamefont {M.}~\bibnamefont {Picco}},\ }\bibfield
  {title} {\bibinfo {title} {{Nonequilibrium critical dynamics of the
  two-dimensional $\ifmmode\pm\else\textpm\fi{}J$ Ising model}},\ }\href
  {https://doi.org/10.1103/PhysRevE.108.064131} {\bibfield  {journal} {\bibinfo
   {journal} {Phys. Rev. E}\ }\textbf {\bibinfo {volume} {108}},\ \bibinfo
  {pages} {064131} (\bibinfo {year} {2023})}\BibitemShut {NoStop}%
\bibitem [{\citenamefont {Tartaglia}\ \emph {et~al.}(2016)\citenamefont
  {Tartaglia}, \citenamefont {Cugliandolo},\ and\ \citenamefont
  {Picco}}]{Tartaglia16}%
  \BibitemOpen
  \bibfield  {author} {\bibinfo {author} {\bibfnamefont {A.}~\bibnamefont
  {Tartaglia}}, \bibinfo {author} {\bibfnamefont {L.~F.}\ \bibnamefont
  {Cugliandolo}},\ and\ \bibinfo {author} {\bibfnamefont {M.}~\bibnamefont
  {Picco}},\ }\bibfield  {title} {\bibinfo {title} {Phase separation and
  critical percolation in bidimensional spin-exchange models},\ }\href
  {https://doi.org/10.1209/0295-5075/116/26001} {\bibfield  {journal} {\bibinfo
   {journal} {Europhys. Lett.}\ }\textbf {\bibinfo {volume} {116}},\ \bibinfo
  {pages} {26001} (\bibinfo {year} {2016})}\BibitemShut {NoStop}%
\bibitem [{\citenamefont {Schramm}(2000)}]{Schramm00}%
  \BibitemOpen
  \bibfield  {author} {\bibinfo {author} {\bibfnamefont {O.}~\bibnamefont
  {Schramm}},\ }\bibfield  {title} {\bibinfo {title} {Scaling limits of
  loop-erased random walks and uniform spanning trees},\ }\href@noop {}
  {\bibfield  {journal} {\bibinfo  {journal} {Isr. J. Math.}\ }\textbf
  {\bibinfo {volume} {118}},\ \bibinfo {pages} {221} (\bibinfo {year}
  {2000})}\BibitemShut {NoStop}%
\bibitem [{\citenamefont {Cardy}(2005)}]{Cardy05}%
  \BibitemOpen
  \bibfield  {author} {\bibinfo {author} {\bibfnamefont {J.}~\bibnamefont
  {Cardy}},\ }\bibfield  {title} {\bibinfo {title} {{SLE for theoretical
  physicists}},\ }\href
  {https://doi.org/https://doi.org/10.1016/j.aop.2005.04.001} {\bibfield
  {journal} {\bibinfo  {journal} {Ann. Phys.}\ }\textbf {\bibinfo {volume}
  {318}},\ \bibinfo {pages} {81} (\bibinfo {year} {2005})},\ \bibinfo {note}
  {special Issue}\BibitemShut {NoStop}%
\bibitem [{\citenamefont {Bauer}\ and\ \citenamefont
  {Bernard}(2006)}]{Bauer06}%
  \BibitemOpen
  \bibfield  {author} {\bibinfo {author} {\bibfnamefont {M.}~\bibnamefont
  {Bauer}}\ and\ \bibinfo {author} {\bibfnamefont {D.}~\bibnamefont
  {Bernard}},\ }\bibfield  {title} {\bibinfo {title} {{2D growth processes: SLE
  and Loewner chains}},\ }\href
  {https://doi.org/https://doi.org/10.1016/j.physrep.2006.06.002} {\bibfield
  {journal} {\bibinfo  {journal} {Phys. Rep.}\ }\textbf {\bibinfo {volume}
  {432}},\ \bibinfo {pages} {115} (\bibinfo {year} {2006})}\BibitemShut
  {NoStop}%
\bibitem [{\citenamefont {Smirnov}(2001)}]{Smirnov01}%
  \BibitemOpen
  \bibfield  {author} {\bibinfo {author} {\bibfnamefont {S.}~\bibnamefont
  {Smirnov}},\ }\bibfield  {title} {\bibinfo {title} {{Critical percolation in
  the plane: conformal invariance, Cardy's formula, scaling limits}},\ }\href
  {https://doi.org/https://doi.org/10.1016/S0764-4442(01)01991-7} {\bibfield
  {journal} {\bibinfo  {journal} {C. R. Acad. Sci. Paris, Ser. I}\ }\textbf
  {\bibinfo {volume} {333}},\ \bibinfo {pages} {239} (\bibinfo {year}
  {2001})}\BibitemShut {NoStop}%
\bibitem [{\citenamefont {Almeida}\ and\ \citenamefont
  {Takeuchi}(2021)}]{Almeida21}%
  \BibitemOpen
  \bibfield  {author} {\bibinfo {author} {\bibfnamefont {R.~A.~L.}\
  \bibnamefont {Almeida}}\ and\ \bibinfo {author} {\bibfnamefont {K.~A.}\
  \bibnamefont {Takeuchi}},\ }\bibfield  {title} {\bibinfo {title}
  {{Phase-ordering kinetics in the Allen-Cahn (Model A) class: Universal
  aspects elucidated by electrically induced transition in liquid crystals}},\
  }\href {https://doi.org/10.1103/PhysRevE.104.054103} {\bibfield  {journal}
  {\bibinfo  {journal} {Phys. Rev. E}\ }\textbf {\bibinfo {volume} {104}},\
  \bibinfo {pages} {054103} (\bibinfo {year} {2021})}\BibitemShut {NoStop}%
\bibitem [{\citenamefont {Joets}\ and\ \citenamefont
  {Ribotta}(1986)}]{Joetz86}%
  \BibitemOpen
  \bibfield  {author} {\bibinfo {author} {\bibfnamefont {A.}~\bibnamefont
  {Joets}}\ and\ \bibinfo {author} {\bibfnamefont {R.}~\bibnamefont
  {Ribotta}},\ }\bibfield  {title} {\bibinfo {title} {{Hydrodynamic transitions
  to chaos in the convection of an anisotropic fluid}},\ }\href
  {https://doi.org/10.1051/jphys:01986004704059500} {\bibfield  {journal}
  {\bibinfo  {journal} {J. Phys. France}\ }\textbf {\bibinfo {volume} {47}},\
  \bibinfo {pages} {595 } (\bibinfo {year} {1986})}\BibitemShut {NoStop}%
\bibitem [{\citenamefont {Kai}\ and\ \citenamefont {Zimmermann}(1989)}]{Kai89}%
  \BibitemOpen
  \bibfield  {author} {\bibinfo {author} {\bibfnamefont {S.}~\bibnamefont
  {Kai}}\ and\ \bibinfo {author} {\bibfnamefont {W.}~\bibnamefont
  {Zimmermann}},\ }\bibfield  {title} {\bibinfo {title} {{Pattern Dynamics in
  the Electrohydrodynamics of Nematic Liquid Crystals}},\ }\href
  {https://doi.org/10.1143/PTPS.99.458} {\bibfield  {journal} {\bibinfo
  {journal} {Prog. Theor. Phys. Supp.}\ }\textbf {\bibinfo {volume} {99}},\
  \bibinfo {pages} {458} (\bibinfo {year} {1989})}\BibitemShut {NoStop}%
\bibitem [{\citenamefont {Hoshen}\ and\ \citenamefont
  {Kopelman}(1976)}]{Hoshen76}%
  \BibitemOpen
  \bibfield  {author} {\bibinfo {author} {\bibfnamefont {J.}~\bibnamefont
  {Hoshen}}\ and\ \bibinfo {author} {\bibfnamefont {R.}~\bibnamefont
  {Kopelman}},\ }\bibfield  {title} {\bibinfo {title} {{Percolation and cluster
  distribution. I. Cluster multiple labeling technique and critical
  concentration algorithm}},\ }\href {https://doi.org/10.1103/PhysRevB.14.3438}
  {\bibfield  {journal} {\bibinfo  {journal} {Phys. Rev. B}\ }\textbf {\bibinfo
  {volume} {14}},\ \bibinfo {pages} {3438} (\bibinfo {year}
  {1976})}\BibitemShut {NoStop}%
\bibitem [{\citenamefont {Saleur}\ and\ \citenamefont
  {Duplantier}(1987)}]{Saleur87}%
  \BibitemOpen
  \bibfield  {author} {\bibinfo {author} {\bibfnamefont {H.}~\bibnamefont
  {Saleur}}\ and\ \bibinfo {author} {\bibfnamefont {B.}~\bibnamefont
  {Duplantier}},\ }\bibfield  {title} {\bibinfo {title} {{Exact Determination
  of the Percolation Hull Exponent in Two Dimensions}},\ }\href
  {https://doi.org/10.1103/PhysRevLett.58.2325} {\bibfield  {journal} {\bibinfo
   {journal} {Phys. Rev. Lett.}\ }\textbf {\bibinfo {volume} {58}},\ \bibinfo
  {pages} {2325} (\bibinfo {year} {1987})}\BibitemShut {NoStop}%
\bibitem [{\citenamefont {Smirnov}\ and\ \citenamefont
  {Werner}(2001)}]{Smirnov01_exponents}%
  \BibitemOpen
  \bibfield  {author} {\bibinfo {author} {\bibfnamefont {S.}~\bibnamefont
  {Smirnov}}\ and\ \bibinfo {author} {\bibfnamefont {W.}~\bibnamefont
  {Werner}},\ }\bibfield  {title} {\bibinfo {title} {Critical exponents for
  two-dimensional percolation},\ }\href
  {https://doi.org/https://dx.doi.org/10.4310/MRL.2001.v8.n6.a4} {\bibfield
  {journal} {\bibinfo  {journal} {Math. Res. Lett.}\ }\textbf {\bibinfo
  {volume} {8}},\ \bibinfo {pages} {729} (\bibinfo {year} {2001})}\BibitemShut
  {NoStop}%
\bibitem [{\citenamefont {Lawler}\ \emph {et~al.}(2002)\citenamefont {Lawler},
  \citenamefont {Schramm},\ and\ \citenamefont {Werner}}]{Lawler02_paper4}%
  \BibitemOpen
  \bibfield  {author} {\bibinfo {author} {\bibfnamefont {G.}~\bibnamefont
  {Lawler}}, \bibinfo {author} {\bibfnamefont {O.}~\bibnamefont {Schramm}},\
  and\ \bibinfo {author} {\bibfnamefont {W.}~\bibnamefont {Werner}},\
  }\bibfield  {title} {\bibinfo {title} {{One-Arm Exponent for Critical 2D
  Percolation}},\ }\href {https://doi.org/10.1214/EJP.v7-101} {\bibfield
  {journal} {\bibinfo  {journal} {Electronic Journal of Probability}\ }\textbf
  {\bibinfo {volume} {7}},\ \bibinfo {pages} {1} (\bibinfo {year}
  {2002})}\BibitemShut {NoStop}%
\bibitem [{\citenamefont {Duplantier}\ and\ \citenamefont
  {Saleur}(1988)}]{Duplantier88}%
  \BibitemOpen
  \bibfield  {author} {\bibinfo {author} {\bibfnamefont {B.}~\bibnamefont
  {Duplantier}}\ and\ \bibinfo {author} {\bibfnamefont {H.}~\bibnamefont
  {Saleur}},\ }\bibfield  {title} {\bibinfo {title} {{Winding-Angle
  Distributions of Two-Dimensional Self-Avoiding Walks from Conformal
  Invariance}},\ }\href {https://doi.org/10.1103/PhysRevLett.60.2343}
  {\bibfield  {journal} {\bibinfo  {journal} {Phys. Rev. Lett.}\ }\textbf
  {\bibinfo {volume} {60}},\ \bibinfo {pages} {2343} (\bibinfo {year}
  {1988})}\BibitemShut {NoStop}%
\bibitem [{\citenamefont {Wieland}\ and\ \citenamefont
  {Wilson}(2003)}]{Wieland03}%
  \BibitemOpen
  \bibfield  {author} {\bibinfo {author} {\bibfnamefont {B.}~\bibnamefont
  {Wieland}}\ and\ \bibinfo {author} {\bibfnamefont {D.~B.}\ \bibnamefont
  {Wilson}},\ }\bibfield  {title} {\bibinfo {title} {{Winding angle variance of
  Fortuin-Kasteleyn contours}},\ }\href
  {https://doi.org/10.1103/PhysRevE.68.056101} {\bibfield  {journal} {\bibinfo
  {journal} {Phys. Rev. E}\ }\textbf {\bibinfo {volume} {68}},\ \bibinfo
  {pages} {056101} (\bibinfo {year} {2003})}\BibitemShut {NoStop}%
\bibitem [{\citenamefont {Beffara}(2008)}]{Beffara08_Hausdorff}%
  \BibitemOpen
  \bibfield  {author} {\bibinfo {author} {\bibfnamefont {V.}~\bibnamefont
  {Beffara}},\ }\bibfield  {title} {\bibinfo {title} {{The dimension of the SLE
  curves}},\ }\href {https://doi.org/10.1214/07-AOP364} {\bibfield  {journal}
  {\bibinfo  {journal} {The Annals of Probability}\ }\textbf {\bibinfo {volume}
  {36}},\ \bibinfo {pages} {1421 } (\bibinfo {year} {2008})}\BibitemShut
  {NoStop}%
\bibitem [{\citenamefont {Blanchard}\ \emph {et~al.}(2017)\citenamefont
  {Blanchard}, \citenamefont {Cugliandolo}, \citenamefont {Picco},\ and\
  \citenamefont {Tartaglia}}]{Blanchard17}%
  \BibitemOpen
  \bibfield  {author} {\bibinfo {author} {\bibfnamefont {T.}~\bibnamefont
  {Blanchard}}, \bibinfo {author} {\bibfnamefont {L.~F.}\ \bibnamefont
  {Cugliandolo}}, \bibinfo {author} {\bibfnamefont {M.}~\bibnamefont {Picco}},\
  and\ \bibinfo {author} {\bibfnamefont {A.}~\bibnamefont {Tartaglia}},\
  }\bibfield  {title} {\bibinfo {title} {{Critical percolation in the dynamics
  of the 2D ferromagnetic Ising model}},\ }\href
  {https://doi.org/10.1088/1742-5468/aa9348} {\bibfield  {journal} {\bibinfo
  {journal} {J. Stat. Mech.}\ }\textbf {\bibinfo {volume} {2017}},\ \bibinfo
  {pages} {113201} (\bibinfo {year} {2017})}\BibitemShut {NoStop}%
\end{thebibliography}
\providecommand{\noopsort}[1]{}\providecommand{\singleletter}[1]{#1}%

\end{document}